\documentclass[10pt,conference]{IEEEtran}
\usepackage[english]{babel} % {english,greek}
\newcommand{\B}[1]{\boldsymbol{#1}}

\usepackage{amsmath}
\usepackage{amssymb}
\usepackage{epsfig}
\usepackage{psfrag}

\begin{document}
\selectlanguage{english}
\title{{Belief Propagation based MIMO Detection Operating on Quantized Channel Output}\vspace{-0.2cm}}
\author{{Amine Mezghani and Josef A. Nossek}
\authorblockA{\\Institute for Circuit Theory and Signal Processing\\ Munich University of
Technology, 80290 Munich, Germany\\
E-Mail: \{Mezghani, Nossek\}@nws.ei.tum.de}\vspace{-0.8cm}}
\maketitle 
\begin{abstract}
In multiple-antenna communications, as bandwidth and modulation order increase, system components must work with demanding tolerances. In particular, high resolution and high sampling rate analog-to-digital converters (ADCs) are often prohibitively challenging to design.  Therefore ADCs for such applications should be low-resolution. This paper provides new insights into the problem of optimal signal detection  based on quantized received signals for  multiple-input multiple-output (MIMO) channels. It capitalizes on previous works \cite{kabashima,tanaka,montanari,guo} which extensively analyzed the unquantized linear vector channel using graphical inference methods. In particular, a ``loopy'' belief propagation-like (BP) MIMO detection algorithm, operating on quantized data with low complexity, is proposed. In addition, we study the impact of finite receiver resolution in fading channels in the large-system limit by means of a state evolution analysis of the BP algorithm, which refers to the limit where the number of transmit and receive antennas go to infinity with a fixed ratio. Simulations show that the theoretical findings might  give accurate results even with moderate number of antennas. \\

%We consider  general multi-antenna fading channels with coarsely quantized outputs, where the channel is %unknown  to the transmitter and receiver. This analysis is of interest in the context of sensor network %communication where  low power and low cost are key requirements (e.g. standard IEEE 802.15.4 %applications). This is also motivated by highly energy constrained communications devices where
%sampling the signal may be more energy consuming than processing or transmitting it. Therefore the %analog-to-digital converters (ADCs) for such applications should be low-resolution, in order to reduce %their cost and power consumption. In this paper, we consider the extreme case of only 1-bit ADC for each %receive signal component. We derive asymptotics of  the mutual information up to the second order in the %signal-to-noise ratio (SNR) under average and peak power constraints   and study the impact of %quantization. We show that up to second order in SNR, the mutual information of a   system with %two-level (sign) output signals incorporates only a power penalty factor of almost $\frac{\pi}{2}$ (1.96 %dB) compared to the system with infinite resolution for all channels of practical interest. This %generalizes a recent result for the coherent case \cite{mezghaniisit2007}. 
\end{abstract}
\section{Introduction}

Most of the contributions on signal detection for multiple-input multiple-output (MIMO) systems assume that the receiver has access to the channel data with infinite precision. In practice, however, a quantizer (A/D-converter) is applied to the received analog  signal, so that the channel measurements can be processed in the digital domain. In ultra-wideband and/or high-speed applications, both the required resolution and speed of the ADCs tend to rise, making them expensive, power intensive and even infeasible \cite{wentzloff}.  This work deals with MIMO channels with quantized outputs, which we will refer to as quantized MIMO systems. 

Recently, graphical inference methods have been applied to the usual unquantized MIMO detection problem.
In \cite{kabashima,tanaka} ``loopy'' BP-like detection algorithms were derived  as low complexity heuristics for computing the marginal distribution of each signal component. Hereby, we provide an extended version of the approximative BP based algorithm relying on the commonly used Gaussian approximation, while taking into account the quantizer operation. Then, we provide a state evolution formalism for analyzing the BP algorithm in the large-system limit, and also several theoretical results on the impact of finite receiver resolution that can be drawn from it. We note that a similar problem has been considered in \cite{nakamura_isita}, however based on the \emph{replica method} from statistical physics \cite{nakamura_isit}. In \cite{guo3}, linear systems with general separable output channels have been considered in the large-system limit.  Although \cite{guo3} could include our quantized MIMO case, only sparse systems have been considered and no simulation results have been provided to validate the theoretical results. In this work, the ``loopy'' BP algorithm operating on quantized dense linear systems is studied theoretically and experimentally. Moreover, our derivation steps for the large-system limit are quite straightforward and well justified.         
%% Although both approach are different, our results coincides with those in . 
 The main advantage of the BP approach compared to \cite{nakamura_isita} is that it is more intuitive and allows to find efficient algorithms and analyze their performance and convergence behavior. 
In order to ease calculations, we restrict
ourselves to real-valued systems. However,
the results can be extended to 
the complex case.
 
%In \cite{mezghaniisit2007,nossek}, we study the effects of quantization from an information theoretical %point of  view for  MIMO systems, where the channel is perfectly known at the receiver. It turns out %that the loss in channel capacity due
%to coarse quantization is surprisingly small at low to moderate SNR. In \cite{mezghaniisit2008} the %block fading SISO noncoherent channel was studied in details. 
% Motivated by these works, we aim to study the communication performance of general noncoherent fading %channels taking into account the coarse quantization. We consider the extreme case of 1-bit quantized %(hard-decision)  MIMO channel with  no CSI at the transmitter and the receiver. 

\par Our paper is organized as follows. Section \ref{section:scmodel} describes the system model. In Section \ref{section:mutual}, an approximative BP-like detection algorithm operating on quantized data is derived; then, we provide a state evolution analysis of the BP algorithm and study the effects of quantization in the large-system limit   in Section \ref{section:mutual2}. Finally, in Section \ref{section:receiver}, some simulation results are presented to numerically validate the theoretical findings.
\label{section:introduction}
%%%%%%%%%%%%%%%%%%%%%%%%%%%%%%%%%%%%%%%%%%%%%%%%%%%%%%%%%%%%%%%%%%%%%%%%%%%%%%%%%%%%%%%%
\section{System Model}
\label{section:scmodel}
We consider a  point-to-point MIMO channel where the transmitter employs $K$ antennas and the  receiver has $N$ antennas. Let the vector $\boldsymbol{x} \in \mathbb{R}^{K}$ comprises the $K$ transmitted i.i.d. symbols, each drawn from a certain distribution $q_0(x)$ with zero mean and variance $c_x$. The unquantized (analog) output vector $\boldsymbol{y}\in\mathbb{R}^{N}$ is related to the input as 
\begin{equation}
 \boldsymbol{y}=\boldsymbol{H}\boldsymbol{x}+\boldsymbol{\eta},
\end{equation}
 where $\boldsymbol{H} \in \mathbb{R}^{N\times K}$ is the channel matrix assumed to be perfectly known at the receiver, and $\boldsymbol{\eta}$ refers to  Gaussian noise vector with covariance $\boldsymbol{R}_{\eta\eta}=\mathrm {E}[\boldsymbol{\eta}\boldsymbol{\eta}^\textrm{T}]=\sigma_0^2{\bf I}$. 

 In a practical system, each receive signal component $y_l$, $1\leq l\leq N$, is quantized by a $b$-bit resolution  scalar quantizer (A/D-converter). Thus, the resulting quantized signals read as 
\begin{equation}
r_{l}={\cal Q}(y_{l}),
\end{equation}
where ${\cal Q}(\cdot)$ denotes the quantization operation.
 For the case that we use a uniform symmetric mid-riser type quantizer \cite{proaksis}, the quantized receive alphabet for each dimension is given by
\begin{equation}
  r_{l}\in \{ (-\frac{2^b}{2}-\frac{1}{2}+k)\Delta;\textrm{ } k=1,\cdots,2^b\}=\mathcal{R},  
\end{equation}    
where $\Delta$ is the quantizer step-size and $b$ the number of quantizer bits, which are set the same for all the quantizers.
%\begin{figure}[h]
%\begin{center}
%\psfrag{H}[c][c]{$\boldsymbol{H}$}
%\psfrag{x}[c][c]{$\boldsymbol{x}$}
%\psfrag{y}[c][c]{$\boldsymbol{y}$}
%\psfrag{r}[c][c]{$\boldsymbol{r}$}
%\psfrag{n}[c][c]{$\boldsymbol{\eta}$}
%\psfrag{M}[c][c]{$K$}
%\psfrag{N}[c][c]{$N$}
%\psfrag{Q}[c][c]{$Q$}
%\psfrag{ML}[c][c]{DEC}
%\psfrag{^x}[c][c]{$\boldsymbol{\hat{x}}$}
%{\epsfig {file=channel_ML.eps, width = 6cm}}
%\caption{Quantized MIMO System.}
%\label{channel_model}
%\end{center}
%\end{figure}\\

With these definitions, the conditional probability distribution of the quantized output given an input $\B{x}$ reads as
\begin{equation}
p_{\B{r}|\B{x}}(\B{r}| \B{x})=\prod_{l=1}^N \rho_0(r_l|\B{h}_l^{\rm T} \B{x}),
\end{equation}
where $\B{h}_l^{\rm T}$ is the $l$-th row of $\B{H}$ and

\begin{equation}
\begin{aligned}
\rho_0(r_l|\B{h}_l^{\rm T} \B{x}) &= \frac{1}{\sqrt{2\pi \sigma_0^2}} \int_{r_{l}^{\rm low }}^{r_{l}^{\rm up }}  {\rm e}^{\frac{(y- \B{h}_l^{\rm T} \B{x}  )^2}{2\sigma_0^2}} dy \\
&= \Phi\left(\frac{r_{l}^{\rm up}-\B{h}_l^{\rm T} \B{x}}{\sigma_0}\right)-\Phi\left(\frac{r_{l}^{\rm low }-\B{h}_l^{\rm T} \B{x}}{\sigma_0}\right)
\end{aligned}
\end{equation}
with $\Phi(x)$ represents the cumulative Gaussian distribution given by
\begin{equation}
\Phi(x)=\frac{1}{\sqrt{2\pi}}\int_{-\infty}^x\exp\left(-\frac{t^2}{2}\right) \rm{d}t.
\end{equation}
Hereby the lower and upper quantization boundaries are
\begin{equation*}
r_{l}^{\rm low }=
\begin{cases}
r_{l}-\frac{\Delta}{2} & \mathrm{for} \quad r_{l}\geq -\frac{\Delta}{2}(2^b-2)\\
-\infty & \mathrm{otherwise,} 
\end{cases}
\end{equation*}
and 
\begin{equation*}
r_{l}^{\rm up }=
\begin{cases}
r_{l}+\frac{\Delta}{2} & \mathrm{for} \quad r_{l}\leq \frac{\Delta}{2}(2^b-2)\\
+\infty & \mathrm{otherwise}.
\end{cases}
\end{equation*}
  
%\cite{ref1}\cite{ref2}\cite{ref3}\cite{wiesel04multiuser}\cite{schubert}\cite{palomar}
%%%%%%%%%%%%%%%%%%%%%%%%%%%%%%%%%%%%%%%%%%%%%%%%%%%%%%%%%%%%%%%%%%%%%%%%%%%%%%%%%%%%%%%%
\section{Approximative BP Detection}
\label{section:mutual}
Our goal is to derive a low complexity detector computing the conditional mean estimate 
\begin{equation}
\hat{\B{x}}={\rm E}[\B{x}|\B{r}],
\end{equation}
based on the knowledge of $q_0(x)$, $\rho_0(\cdot|\cdot)$ and $\B{H}$. This problem is related to the problem of  finding the  marginal probabilities $p_{x_k|\B{r}}(x_k|\B{r})$, for which belief propagation can provide low-complexity approximations. To this end, a factor graph representation is needed.
\subsection{Factor Graph Representation}
\begin{figure}[h]
\centerline{
\psfrag{X1}[c][c]{$x_1$}
\psfrag{X2}[c][c]{$x_2$}
\psfrag{X3}[c][c]{$x_3$}
\psfrag{XK-1}[c][c]{$\!\!\!\!\!\!\!\!\!\!x_{K-1}$}
\psfrag{XK}[c][c]{$x_K$}
\psfrag{Y1}[c][c]{$r_1$}
\psfrag{Y2}[c][c]{$r_2$}
\psfrag{Y3}[c][c]{$r_3$}
\psfrag{Y4}[c][c]{$r_4$}
\psfrag{YN-1}[c][c]{$r_{N-1}$}
\psfrag{YN}[c][c]{$r_N$}
\psfrag{r} [c][c]{$\cdots$}
\psfrag{pi}[c][c]{$~~~~~\pi_{lk}^{t}(x_k)$}
\psfrag{rho}[c][c]{$\!\!\!\!\!\!\!\!\!\!\lambda_{lk}^{t}(x_k)$}
\epsfig{file=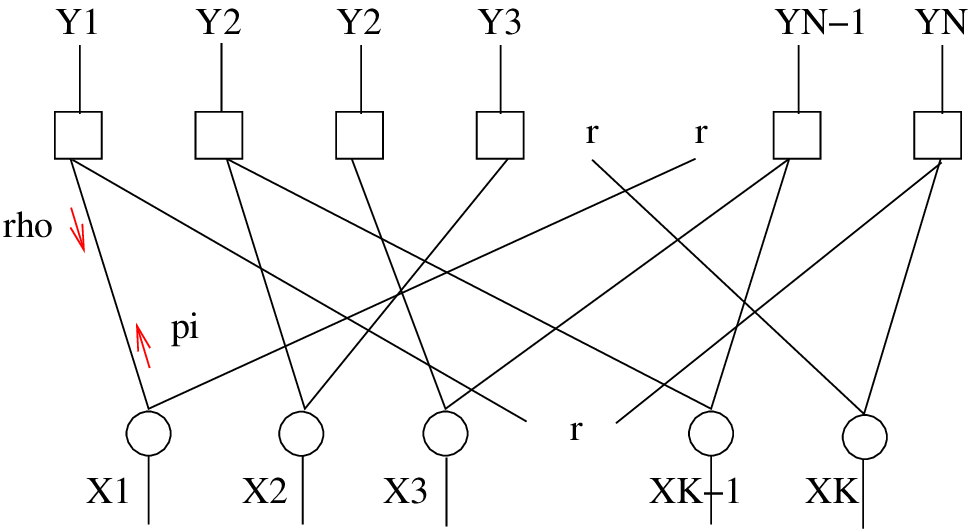, width = 7cm}}
\caption{Factor-graph representation of the quantized MIMO channel.}
\label{fmax}
\end{figure}

In analogy to \cite{guo}, a factor graph representation of the quantized MIMO system is shown in Fig.~\ref{fmax}. Each data stream $x_k$ is represented by a circle, referred to symbol node, and each received quantized signal $r_l$ corresponds to a square, called the signal node. Each edge
connecting $k$ and $l$ represents the corresponding gain
factor $h_{lk}$, if $h_{lk}\neq 0$. Ignoring the cycles in the graph, let us derive the so called   ``loopy'' BP algorithm (or
sum-product algorithm)
from the factor graph representation.\footnote{We note that the BP is optimal for cycle free graphs and performs nearly optimal in sparse graphs. In the case of dense, large enough, channel matrices, it may provide good approximate posteriors as we will see later.} 
%%%%%%%%%%%%%%%%%%%%%%%%%%%%%%%%%%%%%%%%%%%%%%%%%%%%%%%%%%%%%%
\subsection{BP based Detection}
 Each iteration $t$ of BP consists in, first sending  messages from each  signal node $l$ to
each symbol node $k$ (horizontal step), and then vice versa (vertical step). The messages contain the extrinsic information of $x_k$ in form of density functions computed based on the previously received
messages. We denote the symbol-to-signal messages 
 by $\pi_{lk}^{t}(x_k)$ and the signal-to-symbol messages  by $\lambda_{lk}^{t}(x_k)$: \\
%  In each iteration of BP, messages are first sent from symbol nodes to
%signal nodes; each signal node then computes messages to send
%back to the symbol nodes based on the previously received
%messages. These signal-to-symbol messages will then be used
%to generate the new symbol-to-signal messages in the next
%iteration. Let $\pi_{lk}^{t}(x_k)$	
%represent the message from symbol node $k$ to signal node $l$
%and $\lambda_{lk}^{t}(x_k)$
%represent the message in the reverse direction
%at the $t$-th iteration. The messages represents generally the extrinsic information of $x_k$ in some %sense.  The iterative BP algorithm for computing the (approximate) a posteriori distribution of all %symbols is described as follows. 
%$ \textrm{ }$\vspace{0.5cm}\\
%\textit{Input:} Channel state $\B{H}$, output $\B{r}$, prior $q_0(x)$, \\
%\textit{Initialization:} Set t:=0, and \\
%\begin{equation}
%\pi_{lk}^0(x_k)=q_0(x_k),~l=1,\ldots,N,~k=1,\ldots,K. 
%\end{equation}
%\textbf{for} $t=0$ to maximum number of iterations \textbf{do} \\
$\textrm{ }$~~~Horizontal step:
\begin{equation}
\lambda_{lk}^{t}(x_k)=\int \rho_0(r_l|\B{h}_l^{\rm T} \B{x}) \prod_{k' \neq k}[\pi_{lk'}^t(x_{k'}) {\rm d}x_{k'}]
\label{hstep}
\end{equation}
$\textrm{ }$~~~Vertical step:
\begin{equation}
\pi_{lk}^{t+1}(x_k)= \alpha_{lk} q_0(x_k) \prod_{l'\neq l} \lambda_{lk}^{t}(x_k),
\end{equation}
where $\alpha_{lk}$ is a normalization factor so that $\pi_{lk}^{t+1}(x_k)$ is a valid density function. The algorithm is initialized by
\begin{equation}
\pi_{lk}^0(x_k)=q_0(x_k),~l=1,\ldots,N,~k=1,\ldots,K. 
\end{equation}
%\textbf{end} \textbf{for} \\
%\textit{Output:} After convergence is achieved, calculate
%\begin{equation}
%\pi_k(x_k)=\alpha_k q_0(x_k) \prod_{l=1}^{L} \lambda_{lk}(x_k)
%\end{equation}

%%%%%%%%%%%%%%%%%%%%%%%%%%%%%%%%%%%%%%%%%%%%%%%%%%%%%%%%%%%%%%%%%%%%%%%%%%%
%%%%%%%%%%%%%%%%%%%%%%%%%%%%%%%%%%%%%%%%%%%%%%%%%%%%%%%%%%%%%%%%%%%%%%%%%%
\subsection{Approximative BP based Detection Algorithm}
In the case of dense matrices, the complexity of the BP algorithm grows enormously with $K$, since a $(K-1)$-dimensional integration (or summation) has to be performed in the horizontal step. Thus we discuss now an approximation scheme, in analogy to \cite{kabashima} (see also \cite{guo2}), that may be justified in the large-system limit. For this we split the quantity $\B{h}_l^{\rm T}\B{x}$ as
\begin{equation}
   \B{h}_l^{\rm T}\B{x}= \sum_{k'\neq k}h_{lk'}x_{k'} +h_{lk}x_k=w_{lk}+h_{lk}x_k.
   \label{wlk}
\end{equation}
Given the distribution $p_{w_{lk}}\!(w_{lk})$,  (\ref{hstep}) can be rewritten as
\begin{equation} 
 \lambda_{lk}^{t} (x_k) =  \int_{-\infty}^{\infty}\rho_0(r_l|w_{lk}+h_{lk}x_k)p_{w_{lk}}(w_{lk})  {\rm d}w_{lk}.
 \end{equation}
The key remark is that  $w_{lk}$ is a weighted sum of the independent random variables $x_{k'}$  distributed according to $\pi_{lk'}^{t}(x_{k'})$. Due to the central-limit theorem, it can be regarded as a Gaussian random variable in the large-system regime, having the mean and the variance 
\begin{eqnarray}
	\mu_{lk}^{t}&=&\sum_{k' \neq k} h_{lk'} m_{lk'}^{t},  \label {mulk}\\
	C_{lk}^{t}&=& \sum_{k' \neq k} h_{lk'}^2     V_{lk'}^{t}, \label {Clk}
\end{eqnarray}
respectively, where $m_{lk'}^{t}$ and $V_{lk'}^{t}$ are the mean and the variance of $x_{k'}$ according to the distribution  $\pi_{lk'}^{t}(x_{k'})$.  The horizontal step can be now written as follows 
\begin{align}
 \lambda_{lk}^{t} (x_k) &\approx   \int_{-\infty}^{\infty}\rho_0(r_l|w_{lk}+h_{lk}x_k)\frac{{\rm e}^ {-\frac{ (w_{lk}- \mu_{lk}^{t})^2}{2C_{lk}^{t}}}}{\sqrt{2\pi C_{lk}^{t} }}   {\rm d}w_{lk} \nonumber\\
%&= \int_{-\infty}^{\infty}(\Phi(\frac{r_l^{\rm up}-w_{lk}-h_{lk}x_k}{\sigma_0})-\Phi(\frac{r_l^{\rm low}-w_{lk}-h_{lk}x_k}{\sigma_0}))\frac{{\rm e}^ {-\frac{ (w_{kl}- \mu_{lk}^{t})^2}{2C_{lk}^{t}}}}{\sqrt{2\pi C_{lk}^{t} }}   d w_{lk} \\
&= \Phi \Big(\frac{ r_l^{\rm up}\!-\!\mu_{lk}^t \!-\!h_{lk}x_k }{\sqrt{C_{lk}^t+\sigma_0^2}}\Big)\!-\! \Phi \Big(\frac{ r_l^{\rm low}\!-\!\mu_{lk}^t\!-\!h_{lk}x_k  }{\sqrt{C_{lk}^t+\sigma_0^2}}\Big) \nonumber \\
&\dot{=}\rho_{lk}^{t}(r_l|\mu_{lk}^t+h_{lk}x_k). 
\label{rho1}
\end{align}
% The mean $m_{lk}^t$ and the variance $V_{lk}^t$ are computed from $\pi_{lk}^{t} (x_k) $ using %(\ref{mulk}) and (\ref{Clk}), respectively. 

Obviously the approximate BP is a kind of modified parallel interference cancellation (PIC) as in the unquantized case \cite{kabashima}, where  $\mu_{lk}^{t}$ represents an estimate of the interference component in $r_l$ for the stream $k$, and $C_{lk}^{t}$ quantifies its MSE. Even if this approximate BP iteration holds in the limit of infinite number of antennas, it usually  performs well for systems of moderate sizes.

%%%%%%%%%%%%%%%%%%%%%%%%%%%%%%%%%%%%%%%%%%%%%%%%%%%%%%%%%%%%%%%%%%%%%%%%%%%%%%%%%%%%%%%
%%%%%%%%%%%%%%%%%%%%%%%%%%%%%%%%%%%%%%%%%%%%%%%%%%%%%%%%%%%%%%%%%%%%%%%%%%%%%%%%%%%%%%%%
\section{State Evolution Analysis}
\label{section:mutual2}
State evolution analysis (also known as density evolution for the case of sparse matrices) is a powerful tool to study the behavior of belief propagation in the large-system limit \cite{richardson}. The large-system limit means that we consider
the limit when $K$ and $N$ go to infinity, while the ratio $\beta=K/N$ is kept fixed. Under the conjecture that the presented BP based detector  would be asymptotically optimal, this analysis would deliver useful theoretical results about the MIMO system performance under quantization. For the analysis, we assume a random channel matrix $\B{H}$, where the entries $\{h_{l,k}\}$ are i.i.d. with zero mean and variance $1/N$. 
The main idea is to approximate the messages by Gaussian densities, which holds exactly in the large-system limit. In fact, given that $h_{lk} x_k$ in (\ref{rho1}) scales as $1/\sqrt{N}$, it becomes small as $N$ becomes large, and as such we take the second-order expansion of the messages (\ref{rho1}) as
\begin{equation}
\begin{aligned}
 \rho_{lk}^{t} (x_k|\mu_{lk}^t+h_{lk}x_k) \approx &   \rho_{lk}^{t} (x_k|\mu_{lk}^t)  +  \dot{\rho}_{lk}^{t} (x_k|\mu_{lk}^t) h_{lk}x_k+ \\
 & \frac{1}{2}  \ddot{\rho}_{lk}^{t}  (x_k|\mu_{lk}^t)  h_{lk}^2x_k^2+  \mathcal{O}(N^{-3/2}),  
 \end{aligned}
\end{equation}
where $\dot{\rho}_{lk}^{t} (x_k|\mu_{lk}^t)$ and $\ddot{\rho}_{lk}^{t} (x_k|\mu_{lk}^t)$ denote the first order and the second order derivatives of $\rho_{lk}^{t} (x_k|\mu_{lk}^t)  $ with respect to $\mu_{lk}^t$.
Keeping the terms up to the order of $N^{-1}$, and using now the following approximation
\begin{equation}
1+ax+\frac{1}{2}bx^2 = {\rm e}^{ax+\frac{1}{2}(b-a^2)x^2}+\mathcal{O}(x^3),
\end{equation}
the horizontal and vertical steps can be represented as
\begin{equation}
 \lambda_{lk}^{t} (x) \propto  \exp \left[ \theta_{lk}^{t} x- \frac{1}{2}(\theta_{lk}^{t,2} - \Xi_{lk}^{t} )x^2 \right], \textrm{ and}
\end{equation}
\begin{equation}
\pi_{lk}^{t+1} (x) \propto q_0(x) \exp \left[ \sum_{l'\neq l}\theta_{l'k}^{t} x- \frac{1}{2} \sum_{l'\neq l}( \theta_{l'k}^{t,2}-\Xi_{l'k}^{t}) x^2 \right],
\end{equation}
respectively, where we introduced the definitions
\begin{equation}
\theta_{lk}^{t}\dot{=}\frac{ \dot{\rho}_{lk}^{t} (r_l|\mu_{lk}^t) }{ \rho_{lk}^t (r_l|\mu_{lk}^{t}) } h_{lk}  \quad {\rm and} \quad  \Xi_{lk}^{t}\dot{=}\frac{\ddot{ \rho}_{lk}^{t} (r_l|\mu_{lk}^t) }{ \rho_{lk}^{t}(r_l|\mu_{lk}^t) } h_{lk}^2.
\label{thetaxi}
\end{equation}
\par For the following, we assume a random matrix $\B{H}$ with i.i.d. entries and we consider
the large-system limit. The state evolution analysis aims to study the dynamics of the BP detector, i.e. to track the  evolution of the densities parameters, namely $\mu_{lk}^t$, $C_{lk}^t$ from (\ref{mulk}) and (\ref{Clk}), respectively, and additionally
\begin{equation}
z_{lk}^t\dot{=}\sum_{l'\neq l} \theta_{l'k}^{t},~~F_{lk}^t\dot{=}\sum_{l'\neq l} \theta_{l'k}^{t,2},~~G_{lk}^t\dot{=}\sum_{l'\neq l} \Xi_{l'k}^{t}
\label{zlk}
\end{equation}
over the iterations.
Strictly speaking, these variables are random. Relying on a heuristic assumption that the incoming messages $(\theta_{lk}^{t}, \Xi_{lk}^{t})$ to each symbol node remain independent from iteration to iteration, when conditioned on a given transmit symbol $x_k$ and channel $\B{H}$,  and by the central limit theorem, we conclude that  $\mu_{lk}^t$, $C_{lk}^t$ (from (\ref{mulk}) and (\ref{Clk})), $z_{lk}^t$, $F_{lk}^t$ and $G_{lk}^t$ become asymptotically Gaussian. In particular $C_{lk}^t$, $F_{lk}^t$ and $G_{lk}^t$ are sums of terms of order $1/N$, thus they admit asymptotically zero variance. In other words, they become deterministic in the large-system limit at given iteration provided that $\B{H}$ has i.i.d. entries, i.e. independent of the given realization of $\B{H}$,  the received vector $\B{r}$ and the indexes $l$ and $k$; that is
\begin{eqnarray}
 &&\!\!\!\!\!\!\sum_{k' \neq k} h_{lk'}^2     V_{lk'}^{t} \rightarrow \sum_{k' \neq k} {\rm E}_{\B{r}|x_k}[ h_{lk'}^2     V_{lk'}^{t}|\B{H},x_k] \dot{=} C^{t},
 \label{Ct}\\
 &&\!\!\!\!\!\!\sum_{l'\neq l} \theta_{l'k}^{t,2} \rightarrow  \sum_{l'\neq l}  {\rm E}_{\B{r}|x_k}[\theta_{l'k}^{t,2} |\B{H},x_k]\dot{=}F^t,
\label{Ft}\\
&& \!\!\!\!\!\!\sum_{l'\neq l} \Xi_{l'k}^{t} \rightarrow  {\rm E}_{\B{r}|x_k}[ \Xi_{l'k}^{t} |\B{H},x_k] \dot{=} G^t,
\end{eqnarray}
where $\rightarrow$ symbolizes the convergence to the asymptotic limit. Therefore, we can write the messages $\pi^{t+1}_{lk}(x)$ as
\begin{equation}
\begin{aligned}
\pi^{t+1}_{lk} (x) &\rightarrow \alpha q_0(x) \exp \left[ z_{lk}^{t} x- \frac{1}{2} ( F^{t}-G^{t}) x^2 \right].
\end{aligned}
\label{Gt}
\end{equation}
We will return to the calculation of these parameters later on. Now let us consider the joint distribution $p_{w_{lk},\mu_{lk}^t}(w,\mu)$ of the interference term in antenna $l$ for symbol $k$, $w_{lk}$ from (\ref{wlk}), and its estimate $\mu_{lk}^t$ in (\ref{mulk}), given $\B{H}$ and $x_k$.
Again by the central limit argument, $w_{lk}$ and $\mu_{lk}^t$   are asymptotically jointly Gaussian 
\begin{equation}
(w_{lk},\mu^t_{lk}) \sim \mathcal{N}\left(\B{0}, \B{R}_{w,\mu}^t \right),
 \nonumber
\end{equation}
with the covariance matrix $\B{R}_{w,\mu}^t$, having the entries
\begin{equation}
\begin{aligned}
{\rm E}[w_{lk}^2|\B{H},x_k]&=	{\rm E}\Big[\Big(\sum_{k'\neq k} h_{lk'} x_{k'}\Big)^2\Big|\B{H},x_k\Big]\\
&=\sum_{k'\neq k}h_{lk'}^2{\rm E}[ x_{k'}^2|\B{H}] \\
&\rightarrow \beta \int x_{k'} q_0(x_{k'}){\rm d}x_{k'} \dot{=} \beta c_x,
\end{aligned}
\end{equation}
\begin{equation}
\begin{aligned}
{\rm E}[\mu_{lk}^{t,2}|\B{H},x_k]&=	{\rm E}[(\sum_{k'\neq k} h_{lk'} m_{lk'}^t)^2|\B{H},x_k] \\
&= \sum_{k'\neq k}h_{lk'}^2{\rm E}[ m_{lk'}^{t,2}|\B{H}] \\
&\rightarrow \beta {\rm E}_{z_{lk'}^t}\Big[\Big(\int x \pi_{lk'}^{t}(x){\rm d}x\Big)^2\Big] \dot{=} \beta c_{m}^t,
\end{aligned}
\label{cm}
\end{equation}
\begin{align}
{\rm E}[w_{lk}\mu_{lk}^{t}|\B{H},x_k]&= \sum_{k',k''\neq k} h_{lk'}h_{lk''}{\rm E}[ x_{k''}m_{lk'}^{t}|\B{H}] \nonumber \\
&= \sum_{k'\neq k} h_{lk'}^2{\rm E}[ x_{k'}m_{lk'}^{t}|\B{H}] \label{cxm} \\
&\rightarrow \beta {\rm E}_{x_{k'},z_{lk'}^t}\Big[x_{k'}\int x \pi_{lk'}^{t}(x){\rm d}x\Big] \dot{=} \beta c_{x,m}^t.\nonumber 
\end{align}
In the steps above, we used the fact that the expectations with respect to $x_{k'}$ and $z_{lk'}^t$ become asymptotically independent of the indexes $l$ and $k'$. Thus, we obtain the bivariate Gaussian distribution
\begin{equation}
p_{w_{lk},\mu_{lk}^t}(w,\mu)=\frac{{\rm e}^{-\frac{(w-\frac{c_{x,m}^t}{c_m^t}\mu)^2}{2\beta (c_x-\frac{c_{x,m}^{t,2}}{c_m^t}) }}}{\sqrt{2\pi\beta(c_x-\frac{c_{x,m}^{t,2}}{c_m^t})}}      \frac{{\rm e}^{-\frac{\mu^2}{2\beta c_m^t }}}{\sqrt{2\pi\beta c_m^t}}.
\label{pwmu}
\end{equation}
Next, the MSE parameter $C^t$ in (\ref{Ct}) can be expressed as
\begin{align}
\!\! C^t\!  &=\!	\sum_{k'\neq k} {\rm E}\Big[ h_{lk'}^2 \int x^2 \pi_{lk'}^t(x) {\rm d}x - m_{lk'}^{t,2} \Big|\B{H},x_k\Big] \nonumber \\
&= \!\sum_{k'\neq k}h_{lk'}^2 {\rm E}\Big[ \! \int \!\!  x^2 \pi_{lk'}^t(x) {\rm d}x \Big|\B{H},x_k \Big] \! - \!  \sum_{k'\neq k}\! h_{lk'}^2{\rm E}[ m_{lk'}^{t,2}|\B{H}] \nonumber \\
&\rightarrow \! \beta {\rm E}_{x_{k'},z_{lk'}^t}\!\Big[\!\int \! x^2 \pi_{lk'}^{t}(x){\rm d}x\!\Big]-\beta c_{m}^t \dot{=} \beta (c_{\hat{x}}^t-c_{m}^t). \label{cx}
\end{align}
Note that the densities $\rho_{lk}^t(\cdot|\cdot)$ defined for the horizontal step in (\ref{rho1}) become independent of $l$ and $k$, and read as
\begin{equation}
\begin{aligned}
\!\rho^{t}(r|\mu)= \Phi \Big(\frac{ r^{\rm up}-\mu  }{\sqrt{\beta (c_{\hat{x}}^t\!-\!c_{m}^t)\!+\!\sigma_0^2}}\Big)- \Phi \Big(\frac{ r^{\rm low}-\mu  }{\sqrt{\beta (c_{\hat{x}}^t\!-\!c_{m}^t)\!+\!\sigma_0^2}}\Big).
\end{aligned}
\end{equation}
Afterwards, we compute $F^t$ from (\ref{Ft}). For that we need the joint distribution $p_{\B{r},\mu_{lk}^t|x_k}(\B{r},\mu_{lk}^t|x_k)$ for fixed channel $\B{H}$

\begin{align}
&p_{\B{r},\mu_{lk}^t|x_k}(\B{r},\mu_{lk}^t|x_k)= \!\int p_{\B{r},\mu_{lk}^t,w_{lk}|x_k}(\B{r},\mu_{lk}^t,w_{lk}|x_k) {\rm d}w_{lk} \nonumber \\
 &= \! p_{\B{r}_l|x_k} (\B{r}_l|x_k) \int p_{r_l,\mu_{lk}^t,w_{lk}|x_k} (r_l,\mu_{lk}^t,w_{lk}|x_k) {\rm d}w_{lk} \nonumber \\
 &=\!  p_{\B{r}_l|x_k} \!(\B{r}_l|x_k) \!\! \int \!\!\! \rho_0(r_l|w_{lk}\!+\!h_{lk}x_k) p_{w_{lk},\mu_{lk}^t}\!(w_{lk},\mu_{lk}^t)  {\rm d}w_{lk} \nonumber \\
 &\rightarrow \! p_{\B{r}_l|x_k} (\B{r}_l|x_k) \bar{\rho}^{t}(r_l|\mu_{lk}^t+h_{lk}x_k)  \frac{{\rm e}^{-\frac{\mu_{lk}^{t,2}}{2\beta c_m^t }}}{\sqrt{2\pi\beta c_m^t}}, \label{prmu}
\end{align}

\hspace{-0.35cm}where $\B{r}_l$ is the vector containing the elements of $\B{r}$ excluding $r_l$  and the density function $\bar{\rho}^{t}(\cdot|\cdot)$ is obtained using (\ref{pwmu}) by performing the integration as
\begin{equation}
\begin{aligned}
\!\!\bar{\rho}^{t}(r|\mu)\!= \! \Phi \Bigg(\!\frac{ r^{\rm up}-\mu  }{\sqrt{\!\beta(c_x\!-\!\frac{c_{x,m}^{t,2}}{c_m^t})\!+\!\sigma_0^2}}\!\Bigg)\!-\! \Phi \Bigg(\!\frac{ r^{\rm low}-\mu  }{\sqrt{\!\beta(c_x\!-\!\frac{c_{x,m}^{t,2}}{c_m^t})\!+\!\sigma_0^2}}\!\Bigg).
\end{aligned}
\end{equation}
 
From (\ref{Ft}), (\ref{thetaxi}) and (\ref{prmu})  we identify after some straightforward steps the asymptotic non-vanishing term for the parameter $F^t$
\begin{equation}
\begin{aligned}
F^t=\sum_{r\in\mathcal{R}} \int \bar{\rho}^{t}(r|\mu)  \left[ \frac{\dot{\rho}^t(r|\mu)}{\rho^t(r|\mu)}\right]^2 \frac{{\rm e}^{-\frac{\mu^{2}}{2\beta c_m^t }}}{\sqrt{2\pi\beta c_m^t}} {\rm d}\mu, 
\label{Ft2}
\end{aligned}
\end{equation}
where we dropped the indexes $l$ and $k$ due to asymptotic independence in the large-system limit. Similarly, we obtain the parameter $G^t$ in (\ref{Gt}) as follows
\begin{equation}
\begin{aligned}
G^t=\sum_{r\in\mathcal{R}} \int \bar{\rho}^{t}(r|\mu)   \frac{\ddot{\rho}^t(r|\mu)}{\rho^t(r|\mu)} \frac{{\rm e}^{-\frac{\mu^{2}}{2\beta c_m^t }}}{\sqrt{2\pi\beta c_m^t}} {\rm d}\mu. 
\end{aligned}
\end{equation}
We turn now to determine the distribution of $z_{lk}^t$ defined in (\ref{zlk}), which, as mentioned before, follows a Gaussian distribution conditioned on $x_k$. Again by  (\ref{thetaxi}) and (\ref{prmu}), we show that its mean is 
\begin{equation}
\begin{aligned}
&{\rm E}_{\B{r}|x_k,\B{H}}[z_{lk}^t|x_k,\B{H}]=\\
&=\sum_{r\in\mathcal{R}}  \int \sum_{l'\neq l} \bar{\rho}^{t}(r|\mu+h_{l'k}x_k)   \frac{\dot{\rho}^t(r|\mu)}{\rho^t(r|\mu)} h_{l'k} \frac{{\rm e}^{-\frac{\mu^{2}}{2\beta c_m^t }}}{\sqrt{2\pi\beta c_m^t}} {\rm d}\mu \\
&\rightarrow \sum_{r\in\mathcal{R}}  \int  \dot{\bar{\rho}}^{t}(r|\mu)   \frac{\dot{\rho}^t(r|\mu)}{\rho^t(r|\mu)}\frac{{\rm e}^{-\frac{\mu^{2}}{2\beta c_m^t }}}{\sqrt{2\pi\beta c_m^t}} {\rm d}\mu \cdot x_k \dot{=} E^t \cdot x_k, 
\end{aligned}
\end{equation}
and its variance is
\begin{equation}
\begin{aligned}
&{\rm E}_{\B{r}|x_k,\B{H}}[z_{lk}^{t,2}|x_k,\B{H}]-({\rm E}_{\B{r}|x_k,\B{H}}[z_{lk}^t|x_k,\B{H}])^2 \\
&\quad\quad\quad\quad\quad\quad\quad\rightarrow \sum_{l'\neq l}{\rm E}_{\B{r}|x_k,\B{H}}[\theta_{l'k}^{t,2}|x_k,\B{H}] \stackrel{!}{=} F^t.
\end{aligned}
\end{equation}
Note that $z_{lk}^t \approx \sum_{l'=1}^N \theta_{l'k}^t=z_{k}^t$ and thus $\pi_{lk}^t(x_k)\approx \pi_{k}^t(x_k)$ for all $l$ because each $\theta_{lk}$ has a vanishingly small effect on the sum. In summary, we get the conditional density
\begin{equation}
\rho_{G}^t(z_{k}^t|x_k,\B{H})= \frac{1}{\sqrt{2\pi F^t}} {\rm e}^{-\frac{(z_{k}^t-{E^t} x_k)^2}{2{F^t}}}.
\label{dist_z}
\end{equation} 

Since the BP iteration is initialized by $\pi_{lk}(x_k)=q_0(x_k)$, we have from (\ref{cm}), (\ref{cxm}) and (\ref{cx}) the initial parameters $c_m^0=c_{x,m}^0=0$ and $c_{\hat{x}}^0=c_x$, it can be shown by mathematical induction that for all $t$
\begin{equation}
\begin{aligned}
\!\!\!&c_{m}^t=c_{x,m}^t,~~c_{\hat{x}}^t=c_x,~~\bar{\rho}^t(r|\mu) \equiv \rho^t(r|\mu), \\
\!\!\!& E^t=F^t,~~G^t=0,~~\pi_{k}^{t+1}(x_k)\propto q_0(x_k)\cdot \rho_G^t(z_k|x_k) .
\end{aligned}
\end{equation}
Finally, we conclude that the performance of the large-system regime is fully described by the sequential application of two updates for the parameters $F^t$ and $c_m^t$ (cf. (\ref{Ft2}) and  (\ref{cm}))

\begin{equation}
\begin{aligned}
F^t&=\sum_{r\in\mathcal{R}} \int  \frac{\dot{\rho}^t(r|\mu)^2}{\rho^t(r|\mu)} \frac{{\rm e}^{-\frac{\mu^{2}}{2\beta c_m^t }}}{\sqrt{2\pi\beta c_m^t}} {\rm d}\mu,  \\
c^{t+1}_m&=\int \frac{ [\int x \rho_{G}^{t}(z|x)q_0(x)  {\rm d}x  ]^2 }{\int \rho_{G}^{t}(z|x)q_0(x)  {\rm d}x} {\rm d}z,
\end{aligned}
\label{state}
\end{equation}
with the initial state $c_m^0=0$. 
Interestingly, the stationary conditions of the state evolution equations coincides with  the fixed point equation found in \cite{nakamura_isit}  with the replica method. 

%First, we assume qb  qb $ \stackrel{!}{=} $
%k for all . This implies that any
%single chip y has vanishingly small effect in inferring bk. This may
%be justified in the large-system limit, where the number of chips N
%is very large
 
%%%%%%%%%%%%%%%%%%%%%%%%%%%%%%%%%%%%%%%%%%%%%%%%%%%%%%%%%%%%%%%%%%%%%%%%%%%%%%%%%%%%%%%%%
%%%%%%%%%%%%%%%%%%%%%%%%%%%%%%%%%%%%%%%%%%%%%%%%%%%%%%%%%%%%%%%%%%%%%%%%%%%%%%%%%%%%%%%%
\section{Numerical Results}
\label{section:receiver}
\begin{figure}[ht]
\begin{center}
\psfrag{E}[c][c]{$F$}
\psfrag{m}[c][c]{$c_m$}
\psfrag{mtoE}[c][c]{\!\!\!\!\!\!\!\!\!$c_m^t\rightarrow F^t$}
\psfrag{Etom}[c][c]{\!\!\!\!\!\!\!\!\!$F^t\rightarrow c_m^{t+1}$}
{\epsfig {file=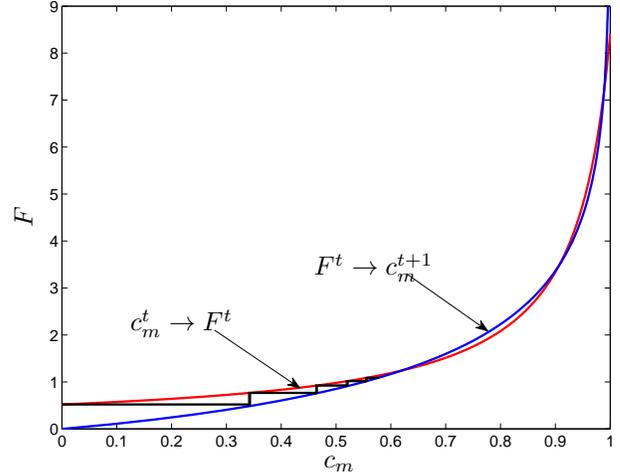, width = 8cm}}
\caption{State evolution chart, $\beta=1.8$, $\sigma_0=0.1$ and $b=4$.}
\label{sim1}
\end{center}
\end{figure}

\begin{figure}[t]
\begin{center}
\psfrag{Pb}[c][c]{BER}
\psfrag{SNR}[c][c]{$10\log_{10}\frac{1}{\sigma_0^2}$}
\psfrag{m}[c][c]{$s$}
\psfrag{mtoE}[c][c]{\!\!\!\!\!\!\!\!\!$s\rightarrow E$}
\psfrag{Etom}[c][c]{\!\!\!\!\!\!\!\!\!$E\rightarrow s$}
{\epsfig {file=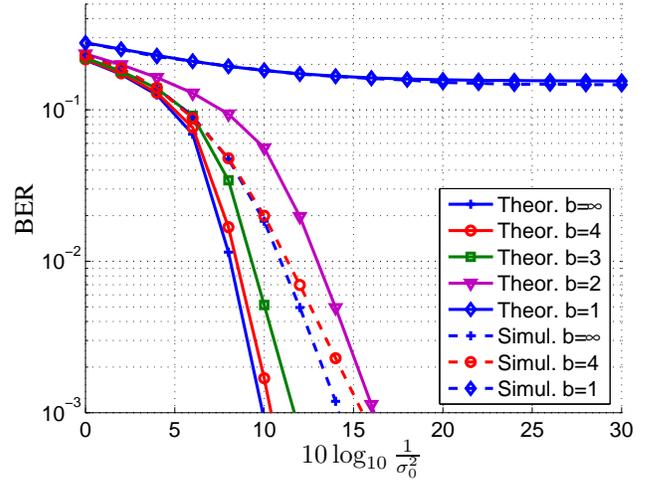, width = 8.5cm}}
\caption{BER for BPSK as  function of $\sigma_0^2$ for $\beta=1$ and different bit resolutions. The simulative results were obtained from the approximative BP detection applied on a $20\times20$ system.\vspace{-0.1cm}}
\label{sim3}
\end{center}
\end{figure}

\begin{figure}[t]
\begin{center}
\psfrag{Pb}[c][c]{BER}
\psfrag{b}[c][c]{$b$}
\psfrag{beta}[c][c]{$\beta$}
\psfrag{mtoE}[c][c]{\!\!\!\!\!\!\!\!\!$s\rightarrow E$}
\psfrag{Etom}[c][c]{\!\!\!\!\!\!\!\!\!$E\rightarrow s$}
{\epsfig {file=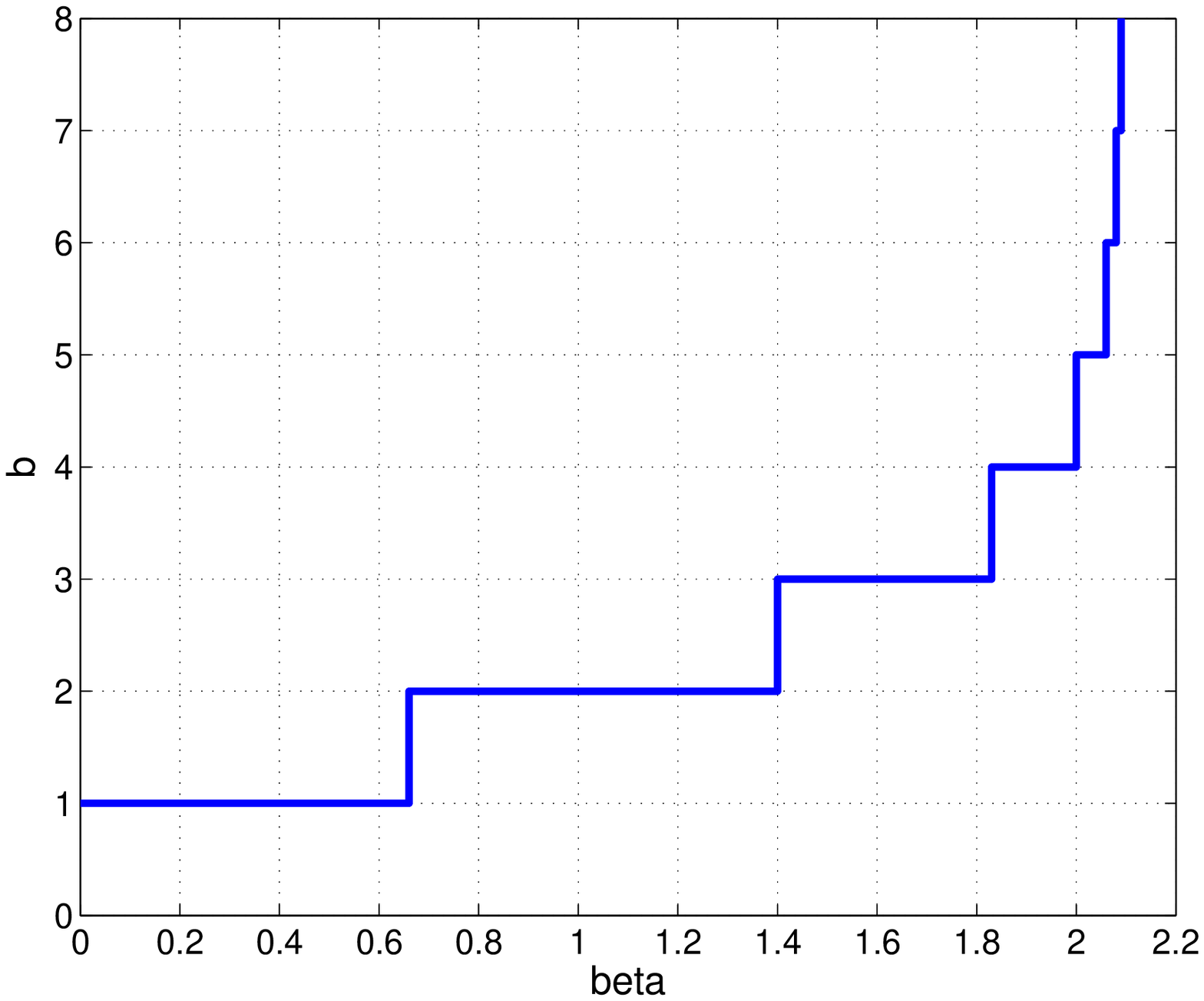, width = 8.7cm}}\vspace{-0.2cm}
\caption{Minimum number of bits needed for BP perfect detection ($P_b\rightarrow 0$ as $\sigma_0\rightarrow0$) as function of $\beta$. }
\label{sim4}
\end{center}
\end{figure}

 Let us consider  BPSK transmission, i.e. $x_k \in \{ -1,+1\}$. We have $c_x=1$ and from (\ref{state}) we get

\begin{align}
  c_m^{t+1}
  &= \int_{-\infty}^{\infty} \frac{1}{2\sqrt{2\pi F^t}} \Big({\rm e}^{-\frac{(z-F^t )^2}{2F^t}}-{\rm e}^{-\frac{(z+F^t )^2}{2F^t}} \Big) \tanh(z) {\rm d}z \nonumber \\
   &= \int_{-\infty}^{\infty} \frac{1}{\sqrt{2\pi F^t}} {\rm e}^{-\frac{(z-F^t )^2}{2F^t}} \tanh(z) {\rm d}z,
 \end{align}
where the second line follows from the antisymmetric property of the $\tanh(z)$  function.
  The equations (\ref{state}) of the state evolution as well as the typical trajectory of the BP iteration are shown in Fig.~\ref{sim1}, for $\beta=1.8$, $\sigma_0=0.1$ and $b=4$. The quantizer step size $\Delta$ has been chosen to minimize the distortion under Gaussian input. We distinguish different fixed points. Clearly, the performance of BP based detection algorithm is characterized by the poor solution as shown in Fig.~\ref{sim1}.    
 Besides, from the distribution of $z_k$ given $x_k$, $\rho_{G}^t(z_k|x_k)$ in (\ref{dist_z}),  which directly affects the decision of $x_k$ through $\pi_k(x_k)$, it immediately follows that the bit error probability (BER) after performing $t$ iterations is given by
  \begin{equation}
  P_b^t= {\rm Pr} (z^t\leq 0)= Q(\sqrt{F^t}),
  \end{equation}
  where $Q(z)=\int_y^{\infty} \frac{1}{\sqrt{2 \pi}}{\rm e}^{-\frac{t^2}{2}} dt$ is the error function.
  The analytical BER performance at the fixed point for $\beta=1$ ($N=K$) is shown in Fig.~\ref{sim3} as function of the SNR$=10\log_{10}\frac{1}{\sigma_0^2}$ for different number of bits. The experimental BERs, carried out for a $20\times20$ system with $b\in\{1,4\}$ and using 10 BP iterations, are also shown for comparison. Obviously, there is a good match between the theoretical results and the Monte Carlo results for $b=1$, while there is a gap for $b=4$ due to the insufficiently high number of antennas. Nevertheless, the analytical curve for $b=4$ still predicts correctly that the performance loss with four bits compared to the ideal case ($b=\infty$) becomes negligible. We note that (\ref{state}) for  the noiseless case ($\sigma_0=0$) always admits a possible  perfect detection solution, i.e. $c_m^\infty=1$ and $F^\infty=\infty$ \cite{nakamura_isita}. However, since the fixed point solution is not unique and BP usually converges to the worst solution, the BER behavior over the SNR might exhibit an error-floor as shown in Fig.~\ref{sim3} for the case $b=1$. The minimum number of bits needed to ensure perfect detection in the noiseless case, i.e. where the recursion  (\ref{state}) evolves to the  unique fixed point solution at $c_m^\infty=1$, is depicted in Fig.~\ref{sim4} as function of the load factor $\beta$. We observe that for a system load $\beta \leq 1$ even 2-bit ADCs might be sufficient for symmetrical systems.
  %%%%%%%%%%%%%%%%%%%%%%%%%%%%%%%%%%%%%%%%%%%%%%%%%%%%%%%%%%%%%%%%%%%%%%
%%%%%%%%%%%%%%%%%%%%%%%%%%%%%%%%%%%%%%%%%%%%
%%%%%%%%%%%%%%%%%%%%%%%%%%%%%%%%%%%%%%%%%%%%%%%
%%%%%%%%%%%%%%%%%%%%%%%%%%%%%%%%%%%%%%%%%%%%%%%%%%%%%%%%%%%%%%%%%%
%%%%%%%%%%%%%%%%%%%%%%%%%%%%%%%%%%%%%%%%%%%%%%%
\vspace{-0.2cm}
%%%%%%%%%%%%%%%%%%%%%%%%%%%%%%%%%%%%%%%%%%%%%%%%%%%%%%%%%%%%%%%%%%
%%%%%%%%%%%%%%%%%%%%%%%%%%%%%%%%%%%%%%%%%%%%%%%%%%%%%%%%%%%%%%%%%%%%%%%%%%%%%%%%%%%%%%%%
%%%%%%%%%%%%%%%%%%%%%%%%%%%%%%%%%%%%%%%%%%%%%%%%%%%%%%%%%%%%%%%%%%%%%%%%%%%%%%%%%%%%%%%
%%%%%%%%%%%%%%%%%%%%%%%%%%%%%%%%%%%%%%%%%%%%%%%%%%%%%%%%%%%%%%%%\%%%%%%%%%%%%%%%%%%%%%%%%
\section{Conclusion}
\label{section:conclusion}
We studied a low complexity detection algorithm based on belief propagation for quantized MIMO systems.  Additionally, a state evolution formalism has been presented to analyze the performance of the BP based detector when operating on quantized data in the large-system limit of fading channels. A set of simulation results was provided, which, in agreement with the analytical results, shows that the BP approach achieves good performance even with low resolution ADCs and moderate number of antennas.  
\bibliographystyle{IEEEbib}
\bibliography{IEEEabrv,references}
\end{document}